\documentclass{article}
\usepackage[utf8]{inputenc}
\usepackage{authblk, quoting, graphicx, xcolor, subcaption}
\usepackage[numbers]{natbib}
\usepackage[bottom]{footmisc}
\pdfoutput=1
\usepackage{hyperref}
\hypersetup{colorlinks,allcolors=blue}

\providecommand{\keywords}[1]{\textbf{Keywords:} #1}

\title{A Case Study in Model Failure? COVID-19 Daily Deaths and ICU Bed Utilisation Predictions in New York State}
\author[1,2]{Vincent Chin}
\author[3]{Noelle I. Samia}
\author[1,2]{Roman Marchant}
\author[4]{Ori Rosen}
\author[5,6,7,8,9,10]{John P.A. Ioannidis}
\author[3]{Martin A. Tanner}
\author[1,2,*]{Sally Cripps}
\affil[1]{ARC Centre for Data Analytics for Resources and Environments, Australia}
\affil[2]{School of Mathematics and Statistics, The University of Sydney, Australia}
\affil[3]{Department of Statistics, Northwestern University, USA}
\affil[4]{Department of Mathematical Sciences, University of Texas at El Paso, USA}
\affil[5]{Stanford Prevention Research Center}
\affil[6]{Department of Medicine, Stanford University, USA}
\affil[7]{Department of Epidemiology and Population Health, Stanford University, USA}
\affil[8]{Department of Biomedical Data Sciences, Stanford University, USA}
\affil[9]{Department of Statistics, Stanford University, USA}
\affil[10]{Meta-Research Innovation Center at Stanford (METRICS), Stanford University, USA}
\affil[*]{Corresponding author: \href{mailto:sally.cripps@sydney.edu.au}{sally.cripps@sydney.edu.au} +61 425-276-967}
\date{\today}

\begin{document}

\maketitle

\begin{abstract}
Forecasting models have been influential in shaping decision-making in the COVID-19 pandemic. However, there is concern that their predictions may have been misleading. Here, we dissect the predictions made by four models for the daily COVID-19 death counts between March 25 and June 5 in New York state, as well as the predictions of ICU bed utilisation made by the influential IHME model. We evaluated the accuracy of the point estimates and the accuracy of the uncertainty estimates of the model predictions. First, we compared the “ground truth” data sources on daily deaths against which these models were trained. Three different data sources were used by these models, and these had substantial differences in recorded daily death counts. Two additional data sources that we examined also provided different death counts per day. For accuracy of prediction, all models fared very poorly. Only 10.2\% of the predictions fell within 10\% of their training ground truth, irrespective of distance into the future. For accurate assessment of uncertainty, only one model matched relatively well the nominal 95\% coverage, but that model did not start predictions until April 16, thus had no impact on early, major decisions. For ICU bed utilisation, the IHME model was highly inaccurate; the point estimates only started to match ground truth after the pandemic wave had started to wane. We conclude that trustworthy models require trustworthy input data to be trained upon. Moreover, models need to be subjected to prespecified real time performance tests, before their results are provided to policy makers and public health officials.
\end{abstract}

\newpage
\section{Introduction}

\begin{quoting}
\noindent
\textit{``I don’t have a crystal ball. Everybody’s entitled to their own opinion, but I don’t operate here on opinion. I operate on facts and on data and on numbers and on projections.''} \cite{herbert} New York Governor Andrew Cuomo - March 24, 2020 \\
\newline
\textit{``Now, people can speculate. People can guess. I think next week, I think two weeks, I think a month, I'm out of that business because we all failed at that business. Right? All the early national experts. Here's my projection model. Here's my projection model. They were all wrong. They were all wrong.''} \cite{cohen} New York Governor Andrew Cuomo - May 25, 2020
\end{quoting}

Forecasting has been very influential in the COVID-19 pandemic. Dealing with a new virus and with a lot of uncertainties surrounding its eventual impact, policy makers have widely used and depended upon predictions made by various models. These predictions refer to critical issues such as the number of anticipated deaths with and without different interventions and the number of hospital beds, ICU beds, and ventilators that would be needed to deal with the surge of the epidemic waves. There is concern that while these models are useful, they can also be very misleading (\cite{nick}, \cite{holm}, \cite{ji}). It is important to understand their performance and their limitations and to try to learn from their failures. This may help generate some better standards for the construction, validation, and use of these models. 

In this article, we evaluate four models for predicting the daily death counts attributable to COVID-19 for the period March 25 to June 5 for the state of New York (NY), as well as one early model that predicted ICU bed utilisation in NY.  The models evaluated are those constructed by the Institute of Health Metrics and Evaluations (IHME) \cite{covid2020forecasting}, Youyang Gu (YYG) \cite{yyg}, the University of Texas at Austin (UT) \cite{ut}, and the Los Alamos National Laboratory (LANL) \cite{lanl}.  These models were chosen because they provide daily death count predictions, as well as 95\% prediction intervals (PIs) for each prediction. The IHME model began producing forecasts from March 25, the corresponding dates for YYG, UT and LANL are April 2, April 14 and April 16, respectively.  We evaluate these models based on two criteria. The first criterion is the accuracy of the point estimates and the second criterion is the accuracy of the uncertainty estimates of those predictions.  With regard to accuracy of prediction, we do not find a model that distinguishes itself from the pack. Most concerning, across models only 10.2\% of the predictions fall within 10\% of their training ground truth, irrespective of distance into the future.  For accurate assessment of uncertainty, the LANL model had observed coverage most closely matching the nominal 95\% coverage. Unfortunately, the LANL model did not commence predictions until April 16, approximately a month \textbf{after} the final US state declared a state of emergency and eleven days \textbf{after} the final US state entered lock-down, thus it played no role in the initial major decisions made by key policy makers in NY, as well as Washington DC.   Regarding the prediction of ICU bed utilisation, the single model (IHME) was highly inaccurate and the point estimates only started to match ground truth by early May, after the pandemic wave had started to wane. Two major takeaways from this research are that 
\begin{enumerate}
    \item \textbf{Serious thought and investment should be made in quality data collection when it comes to COVID-19 daily death data, as well as COVID-19 resource utilisation}.
    \item \textbf{Models need to be subjected to real time performance tests, before their results are provided to policy makers and public health officials. In this paper, we provide examples of such tests, but irrespective of which tests are adopted, they need to be specified in advance, as one would do in a well-run clinical trial}.
\end{enumerate}

\section{The Data}

In order to evaluate the models, it is necessary to define the actual ground truth number of daily deaths.  This task is more problematic than it would first appear, as there is no one source of ground truth. The models YYG and LANL use the raw daily death counts in NY reported by the Johns Hopkins University (JHURD) \cite{jhurd} for training, IHME uses daily deaths reported by the New York Times (NYT) \cite{nyt}, while UT uses NYT data until May 5 to train their model before switching to another version of the JHU data which is known as the JHU time series (JHUTS) data \cite{jhuts}. The JHUTS data is an update of the JHURD to correct for reporting errors. Many modellers (e.g. \cite{yyg}) view the JHURD data as the gold standard, though \cite{lanl} raise concerns with the JHURD data, as well.

\begin{figure}[b!]
    \centering
    \includegraphics[clip,width=.95\textwidth]{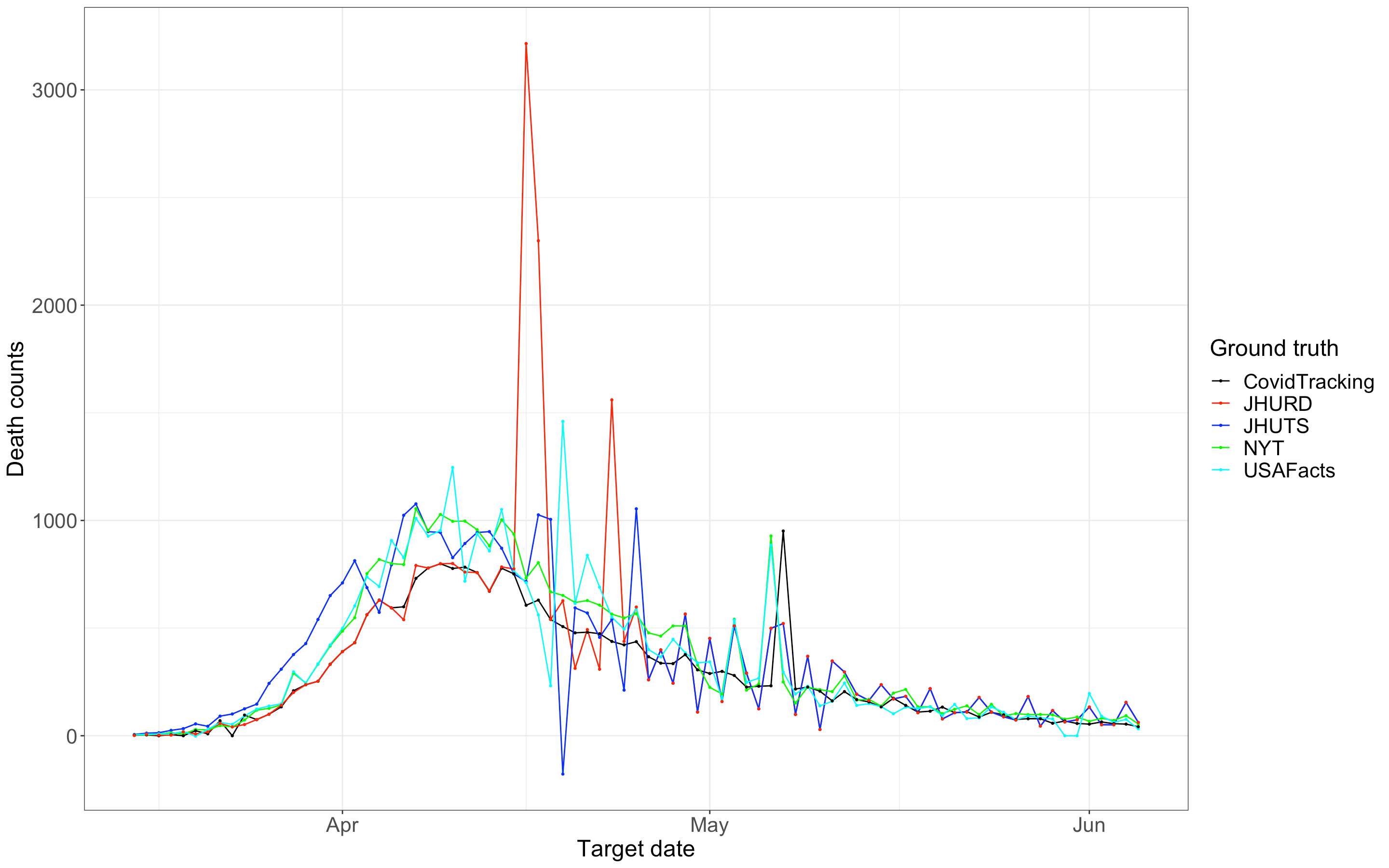}
    \caption{A comparison of the daily death counts ground truth from CovidTracking (black), JHURD (red), JHUTS (dark blue), NYT (green) and USAFacts (light blue) for the period March 15 to June 5 for NY.}
    \label{fig:gt comparison}
\end{figure}
Figure~\ref{fig:gt comparison} presents the ground truth data for the state of NY as reported by JHURD (red), JHUTS (dark blue), NYT (green), as well as two additional sources: CovidTracking \cite{covidtracking} (black) and USAFacts \cite{usafacts} (light blue), from March 15 to June 5. The point of this figure is to demonstrate that these ground truths can vary substantially from each other and have features which are artefacts of the way in which deaths are reported.  Of particular note is the early lag between JHURD and JHUTS, as well as the more smoothed process presented by the NYT data.  It is noted that the NYT data are the confirmed COVID-19 cases from January 22 to May 6, while from May 7 on-wards, the NYT data include the confirmed and probable cases using criteria that were developed by local and states government (\cite {git1}).  Both JHURD and JHUTS include confirmed and probable cases. In all sources, the actual ground truth number of daily deaths is calculated by taking the difference of cumulative deaths.

Figure~\ref{fig:gt comparison} also shows evidence of large swings in the number of reported daily deaths, again probably due to lags or corrections in reporting.  Indeed, the JHUTS data show a negative value for the number of deaths in NY on April 19.  This negative value is, in turn, a result of updates to the JHURD to correct reporting errors (\cite {git2}). This value is clearly incorrect. However, without information regarding the details of cases at the individual level, it is not possible to correct these data.  Any attempt to smooth the data raises the question of how the choice of smoothing technique may affect any conclusions drawn from the data and \cite{marchant2020learning} raise numerous concerns in this regard.  In summary, coding, counting and reporting COVID-19 deaths is highly complex and is beyond the scope of this paper (see e.g. \cite{pappas}).  Accordingly, we have chosen to work with the data provided by these sources and will evaluate each model according to the data the developers have chosen to use for training purposes, as well as to all three ground truths, namely, NYT, JHURD and JHUTS.

\section{Accuracy of the Point Estimates}

\begin{figure}[b!]
    \centering
    \includegraphics[clip,width=.95\textwidth]{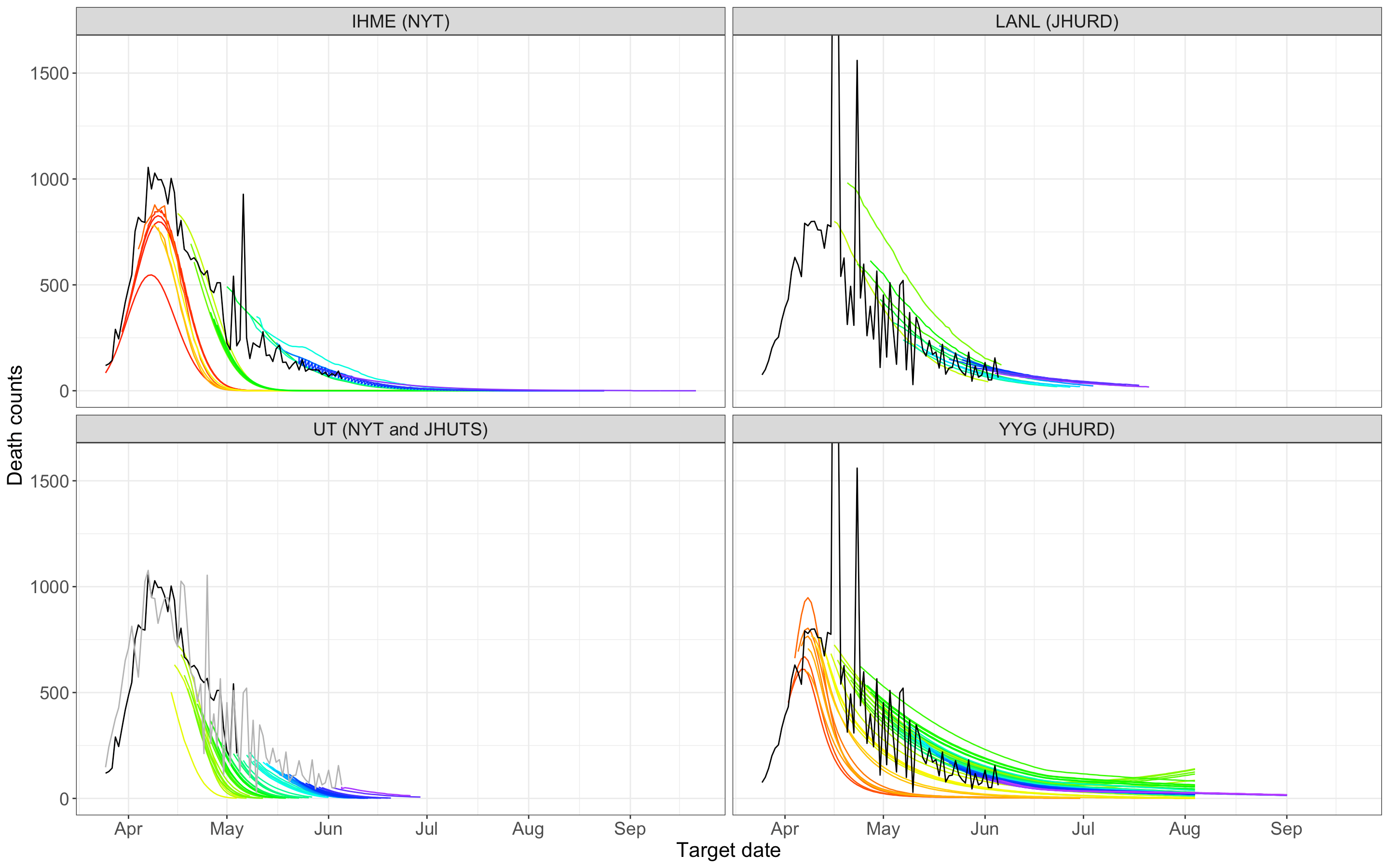}
    \caption{The forecast time series made by each model, along with the ground truth (black) used to train each model. The UT model uses the NYT data (black) until May 5 before switching to the JHUTS data (grey), whereby the negative value for the daily deaths on April 19 (see Figure~\ref{fig:gt comparison}) is removed before the model is trained.}
    \label{fig:spaghetti}
\end{figure}
Figure~\ref{fig:spaghetti} shows the actual data used to develop each of the models, as well as the time series of forecasts made by each of the models. One can see the spike in the number of deaths reported by the NYT in early May following the inclusion of probable cases.  
Figure~\ref{fig:spaghetti} displays only the point estimates of the forecasts.  The time series of forecasts are colour-coded such that the earliest/latest forecasts are at the red/blue end of the colour spectrum. For example, the deepest red curves are the time series of forecasts made in late March, the yellow curves are time series of forecasts made in early April and so on, until the violet curves which represent the most recent time series of forecasts.

\begin{figure}[t!]
    \centering
    \includegraphics[clip,width=\textwidth]{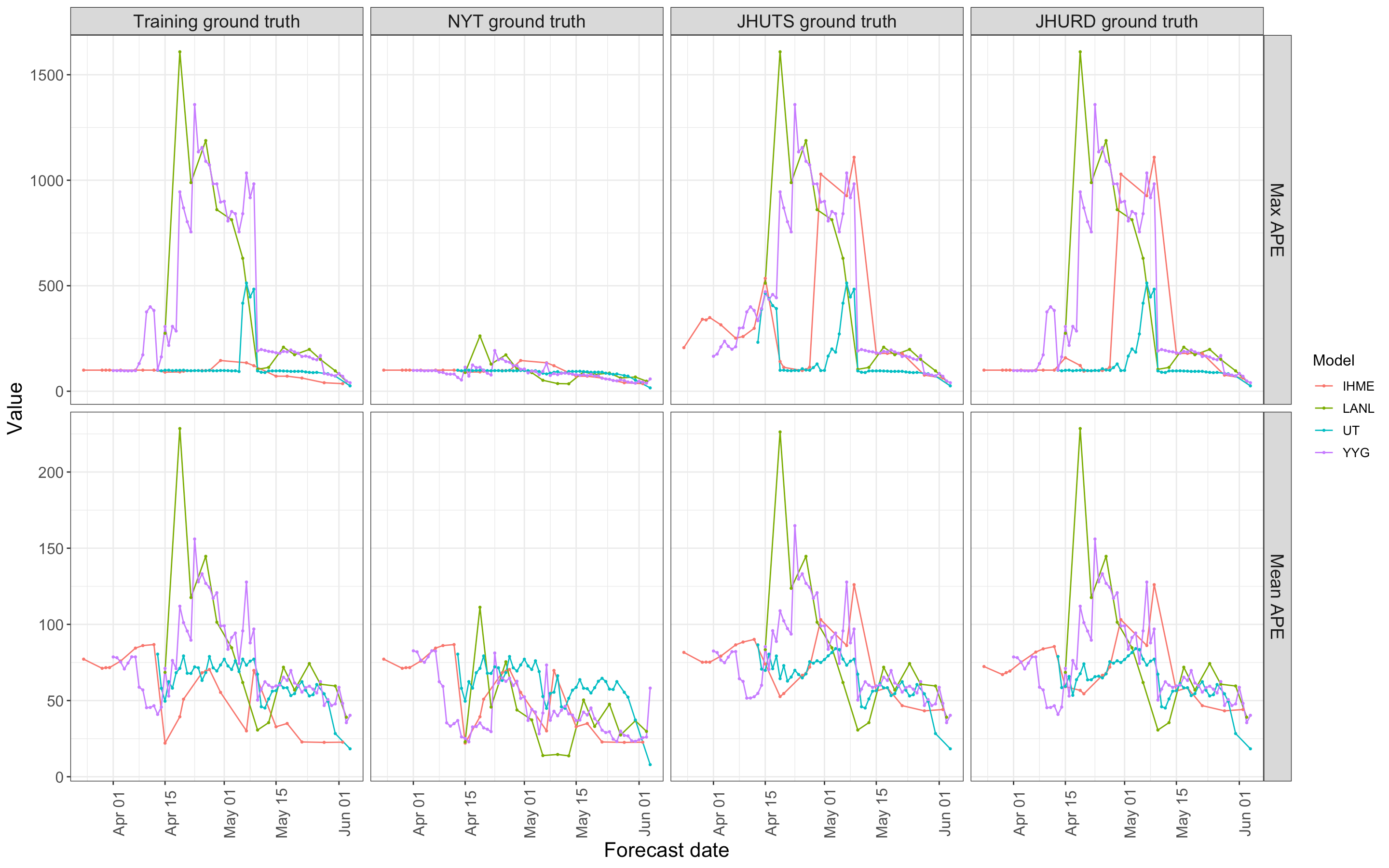}
   
    \caption{Discrepancies between each model and the ground truth, as measured by the maximum absolute percentage error (top) and the mean absolute percentage error (bottom), for each version of the ground truth.}
    \label{fig:performance}
\end{figure}

To evaluate each of the forecast time series in Figure~\ref{fig:spaghetti}, we computed two metrics for each forecast. These metrics are the mean absolute percentage error and the maximum absolute percentage error, whereby the percentage error is computed from {\it (ground truth - predicted value)/(ground truth)$\times$100\%}. 
For the mean absolute percentage error, the percentage of discrepancy between the given model’s prediction and the ground truth was computed for each model for each day for each time series.  This information was then averaged over the entire duration of the forecast for a particular time series. The maximum absolute percentage error was computed by taking the maximum of the absolute percentage errors for each forecast and for each model.  For example, the first forecast made by IHME was on March 25 and we compare the forecast time series for the period from March 25 until June 5 with the ground truth time series over that same period by calculating the two metrics discussed above.  We then repeat the process for each date a forecast time series was issued, and for each of the models. To make a fair comparison on dates where the forecast time series was made by at least two of the models, we truncate the forecast time series at the last prediction date of the shortest time series. These values are plotted over time in Figure~\ref{fig:performance} for each version of the ground truth.

In Figure~\ref{fig:performance}, we see that while some models may perform better or worse over subsets of the time frame of interest, no one model clearly dominates throughout with respect to either of the metrics for any version of the ground truth data.


\section{Accuracy of Uncertainty Quantification} \label{sec:uq}

We now turn to the subject of uncertainty quantification of these models.  Each of the models provides estimates of uncertainty, where the IHME, YYG and LANL forecasts give 95\% PIs, while the UT provides 90\% PIs for predictions made prior to the forecast date of May 16. To translate these 90\% PIs to 95\% PIs for the UT model, we take the log of the prediction and the 90\% PIs; calculate the difference between the log of the prediction and the log of the 90\% PI limits and multiply this difference by a factor of 1.96/1.64. We recompute the 95\% PIs on the log scale before transforming them back to the original scale.

\begin{figure}[h!]
    \begin{subfigure}{\textwidth}
    \centering
    \includegraphics[clip,width=\textwidth]{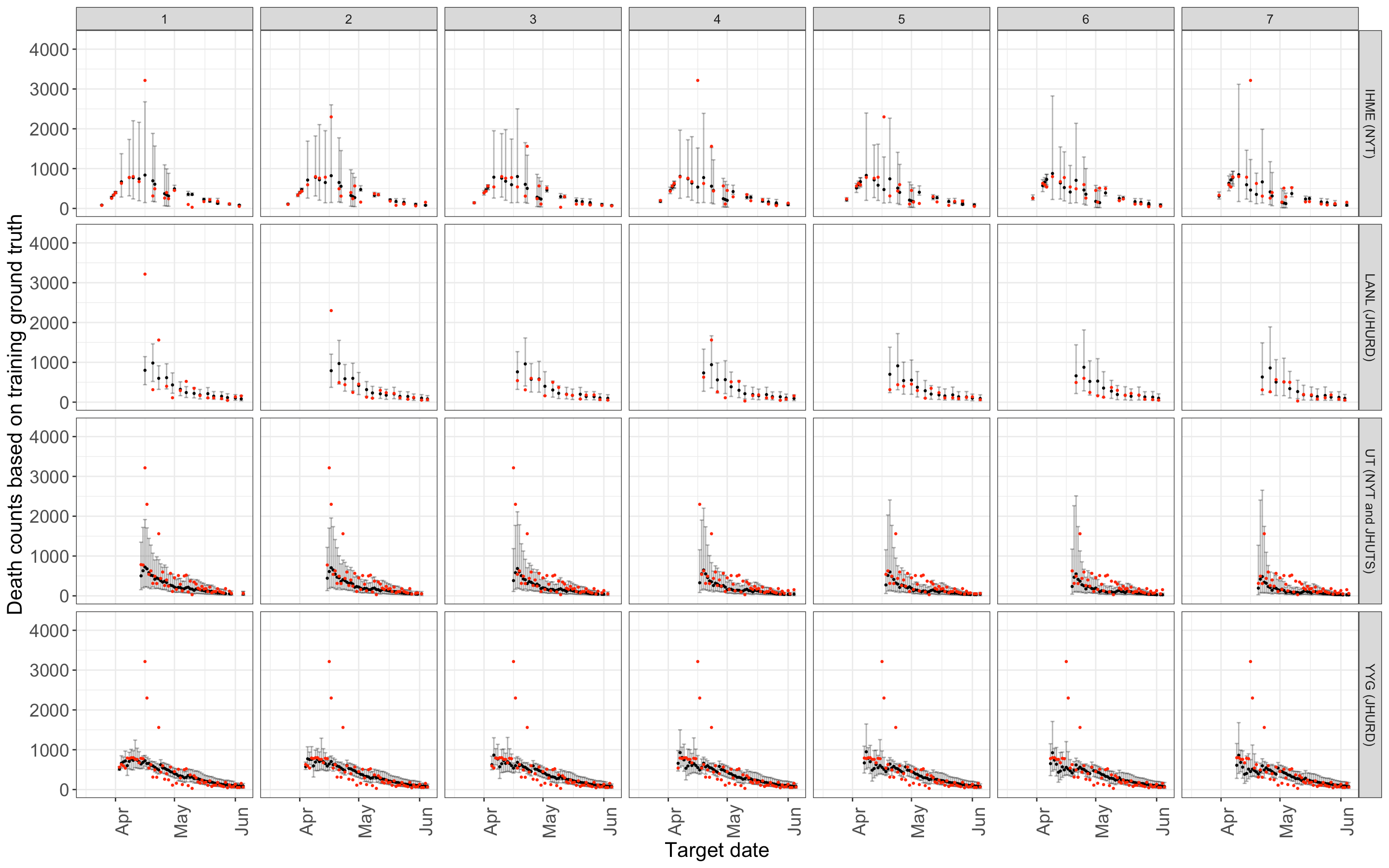}
    \caption{1- to 7-step-ahead predictions.}
    \label{fig:1-7 step}
    \vspace{3em}
    \end{subfigure}
    \begin{subfigure}{\textwidth}
    \centering
    \includegraphics[clip,width=\textwidth]{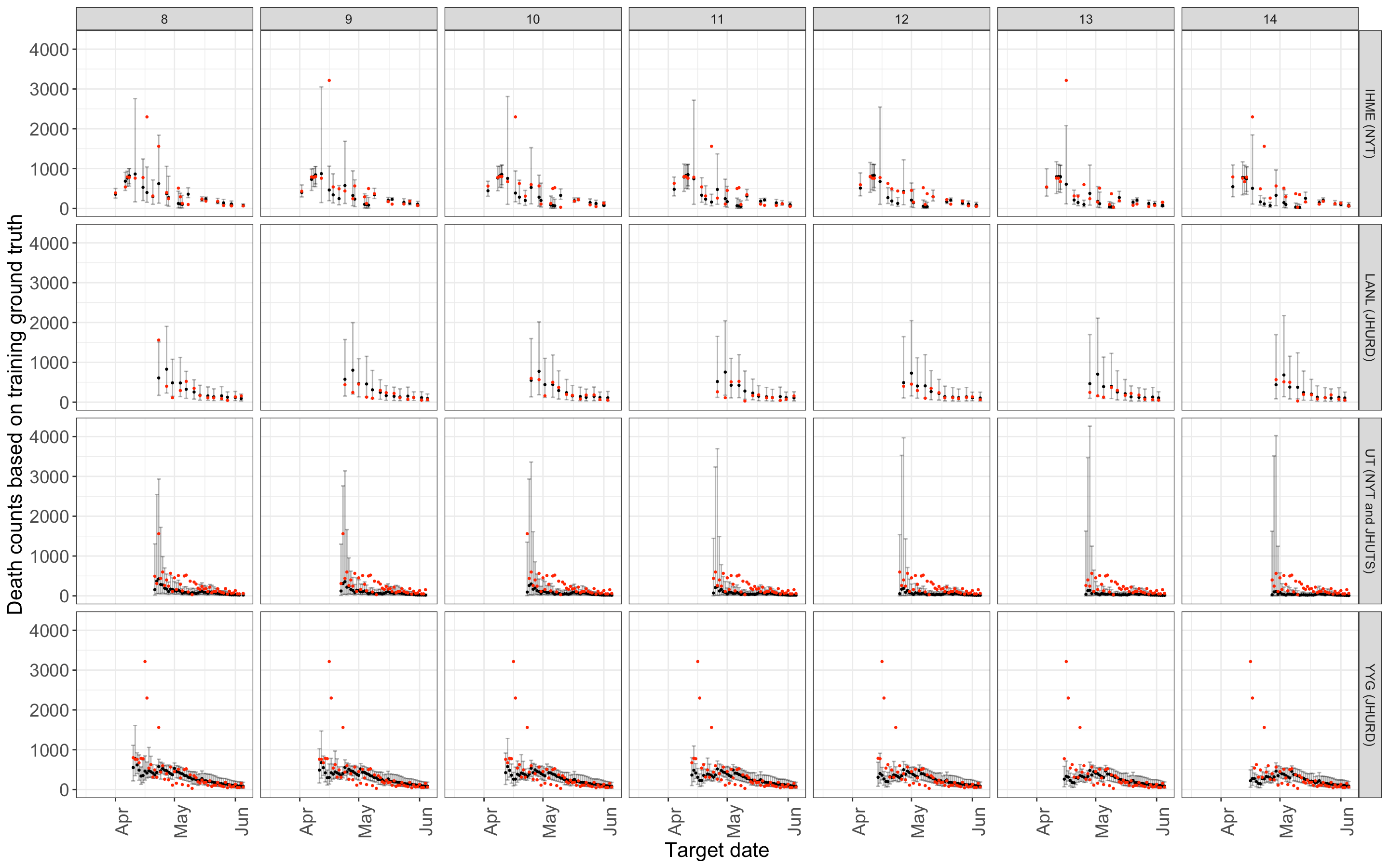}
    \caption{8- to 14-step-ahead prediction}
    \label{fig:8-14 step}
    \end{subfigure}
    \caption{Different step-ahead predictions (black dots) by each model and their 95\% PIs (gray bars), along with the ground truth (red dots) used to train each model.}
    \label{fig:95pi}
\end{figure}

Figure~\ref{fig:95pi} presents plots of the 95\% PIs for various predictions made by the models and the training ground truth. We follow \cite{marchant2020learning} and define a $k$-\textit{step-ahead} prediction and PI for a particular date, to be the prediction and accompanying PI made $k$ days in advance of that date.  For example, for June 3, a 1-step-ahead prediction and PI are the prediction and accompanying interval for June 3 made on the forecast date of June 2, while the 2-step-ahead prediction and PI for June 3 would be made on the forecast date of June 1, etc.  The columns of Figure~\ref{fig:95pi} relate to the number of step-ahead predictions ranging from 1 to 7 in panel \ref{fig:1-7 step} and from 8 to 14 in panel \ref{fig:8-14 step}, while the rows of Figure~\ref{fig:95pi} correspond to the different models.

Figure~\ref{fig:95pi} shows a number of interesting features.  First, as documented in \cite{marchant2020learning}, the IHME model undergoes a number of dramatic changes in the calculation of the prediction and the corresponding PIs.  The original IHME model underestimates uncertainty and 45.7\% of the predictions (over 1- to 14-step-ahead predictions) made over the period March 24 to March 31 are outside the 95\% PIs.  The IHME model was revised on April 2 and made no predictions on April 1 and April 2. In the revised model, for forecasts from of April 3 to May 3 the uncertainty bounds are enlarged, and most predictions (74.0\%) are within the 95\% PIs, which is not surprising given the PIs are in the order of 300 to 2\,000 daily deaths.
Yet, even with this major revision, the claimed nominal coverage of 95\% well exceeds the actual coverage. On May 4, the IHME model undergoes another major revision, and the uncertainty is again dramatically reduced with the result that 47.4\% of the actual daily deaths fall outside the 95\% PIs -- well beyond the claimed 5\% nominal value. It is concerning, nevertheless, that the uncertainty estimates of the IHME model seem to improve with the forecast horizon, so that for the original model and latest IHME model update, more observed values fall within the 95\% PIs for the 7-step-ahead prediction than for the 1-step-ahead prediction.

Second, Figure~\ref{fig:95pi} shows that the YYG model does not perform well in terms of uncertainty quantification, as there are many more actual deaths lying outside the 95\% PIs than would be expected.  Taken across the entire time period, the proportion of actual deaths lying outside the 95\% PIs is 31.1\%.  We do, however, note that this percentage improves over time.  From the forecast date of May 1 on-wards, the fraction of actual deaths lying outside the 95\% PIs ranges from 28.6\% for the 1-step-ahead prediction to 13.6\% for the 14-step-ahead prediction, in comparison to 44.8\% for the 1-step-ahead prediction to 48.3\% for the 14-step-ahead prediction prior to this date.

\begin{figure}[b!]
    \centering
    \includegraphics[clip,width=\textwidth]{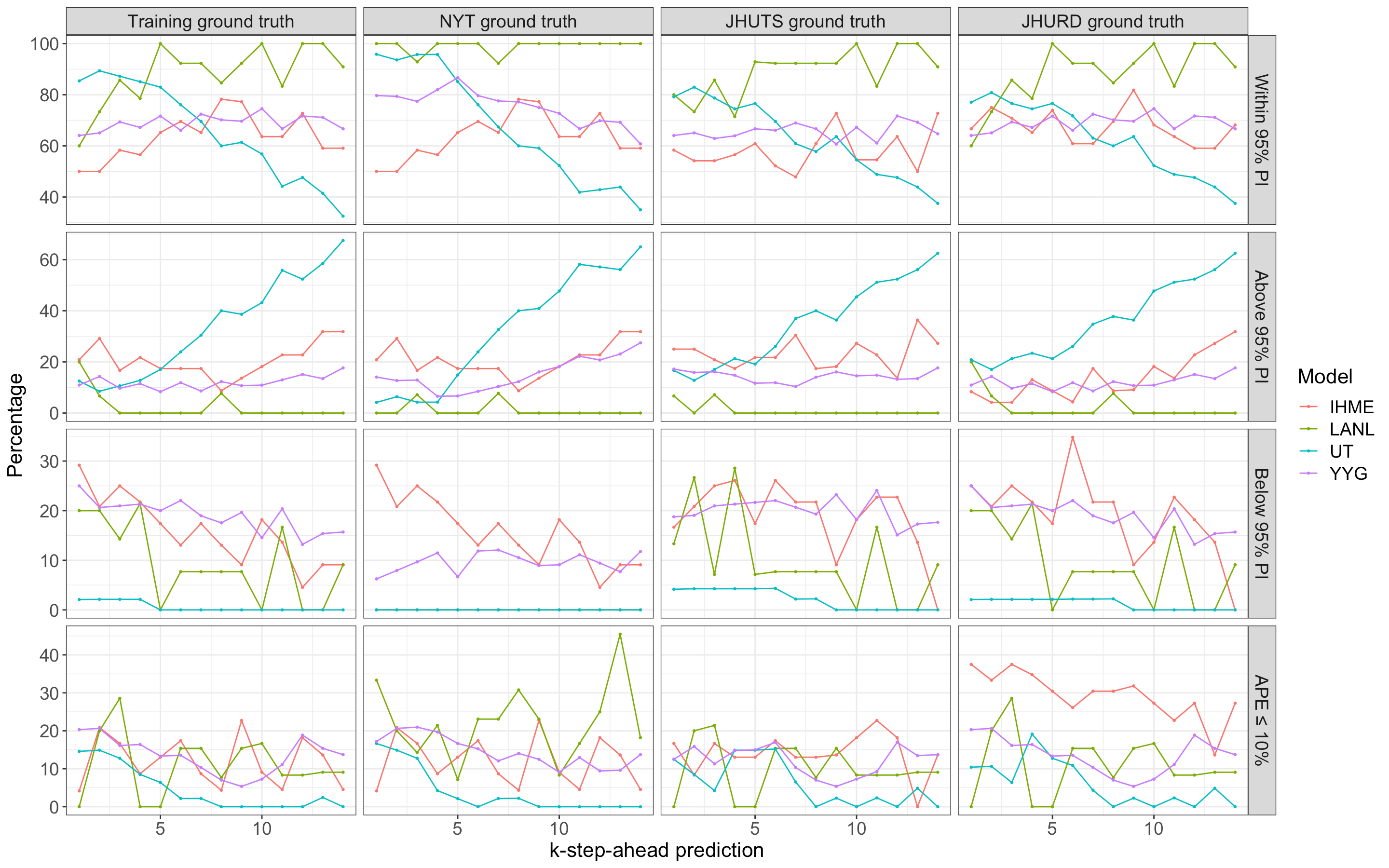}
    \caption{Percentage of the number of daily deaths within, above and below the $k$-step-ahead 95\% PIs. The last panel shows the percentage of $k$-step-ahead predictions which fall within 10\% of the ground truth.}
    \label{fig:percentage 95pi}
\end{figure}

Similarly, regarding the UT and LANL forecasts, neither has observed coverage consistently matching the 95\% nominal coverage as shown in Figure~\ref{fig:percentage 95pi} (first plot in the top panel).  For the UT model, the fraction of actual deaths lying outside the 95\% PIs ranges from 14.6\% for the 1-step-ahead prediction to 67.5\% for the 14-step-ahead. The corresponding figures for the LANL are 40.0\% for the 1-step-ahead prediction to 9.1\% for the 14-step-ahead. 

The second, third and fourth plots in the top panel (one for each version of ground truth) in Figure~\ref{fig:percentage 95pi} present this information from a slightly different perspective.  In particular, from this point of view, for the 5-14 step-ahead predictions, LANL had the best observed coverage compared to the nominal 95\% level, with short-term predictions tending to overestimate the ground truth.  The PIs for the UT model are seen to deteriorate for predictions out into the future, with a tendency of the model to underestimate the daily number of deaths.  The remaining two models, YYG and IHME, tended to provide daily death prediction PIs that systematically miss the nominal 95\% coverage level, irrespective of distance into the future. 

Examining the last panel in Figure~\ref{fig:percentage 95pi}, with the exceptions of IHME evaluated on JHURD and LANL evaluated on the NYT ground truth, no model had more than 30\% of daily death predictions falling within 10\% of the ground truth, with the UT having virtually no predictions within a 10\% bound of the ground truth, out into the future.  When evaluated on their training ground truth, only 10.2\% of the predictions  fall within 10\% of their training ground truth, irrespective of distance into the future.

\section{Prediction of ICU Bed Utilisation}

We now turn our attention to ICU bed utilisation in New York State.  The only model that provides early daily predictions and PIs for NY ICU bed usage is the IHME model and we limit our attention to this model. The IHME model was very influential in early decision-making at the highest levels of the United States government, in regard to the allocation of resources for ICU bed usage, having been mentioned at White House Press conferences, including March 31, 2020 \cite{whitehouse}.

\begin{figure}[b!]
    \centering
    \includegraphics[clip,width=\textwidth]{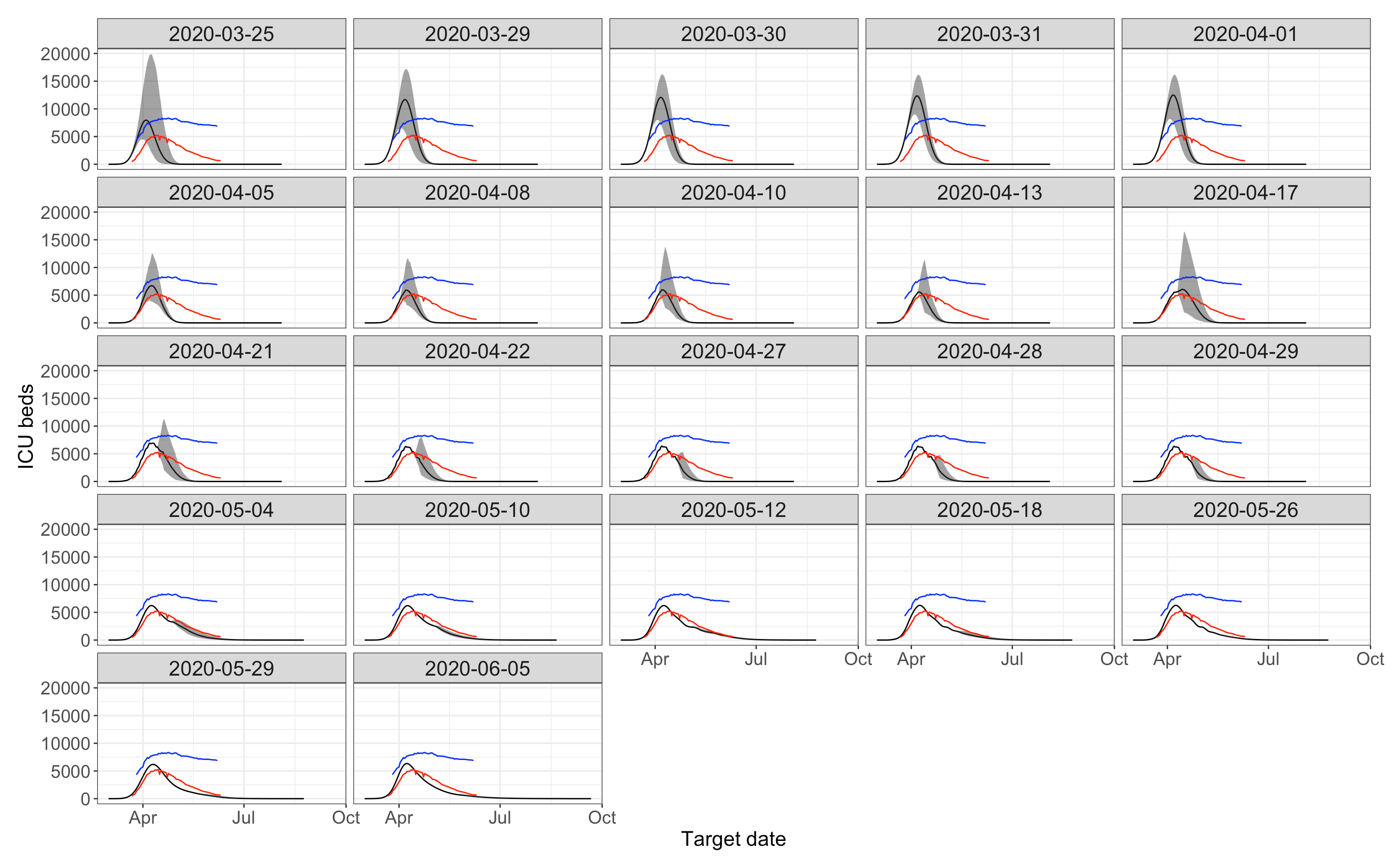}
    \caption{Predicted ICU bed usage (black) and its 95\% PIs (grey shaded area) in NY for each reporting date, along with the ground truth (red) and the maximum ICU capacity inclusive of non-COVID-19 ICU beds (blue) obtained from THE CITY.}
    \label{fig:icu}
\end{figure}

Figure~\ref{fig:icu} presents the IHME estimates (black) and 95\% PIs (grey) for ICU bed usage in NY, together with the ground truth (red) and the maximum ICU capacity (blue; inclusive of non-COVID-19 ICU beds) obtained from THE CITY \cite{thecity}. Each subplot in the figure corresponds to a day for which a prediction was made.  For example, the first subplot is for the prediction made by IHME on March 25, the second for the prediction made on March 29, and so on until the last prediction made on June 5.  The prediction intervals start at the date for which the prediction was made, and thus the gray shaded area, which represents the PIs, shifts to the right for subsequent subplots.  

Figure~\ref{fig:icu} shows that the early forecasts of ICU bed utilisation were highly inaccurate--the prediction intervals for ICU beds made on March 25 through March 31 for the ICU bed usage on April 1 did not contain the actual value despite the width of these PIs being in the order of 5\,000 to 15\,000.  Over this period, the model seriously over-predicted the ICU bed usage. However, by the third week in April through early June, the point predictions of the IHME model systematically \textbf{underestimated} ICU bed utilisation. In fact, in late April, the model predicts zero bed usage by mid-May.

\section{Conclusion}
In a major crisis like COVID-19, policy makers and public health officials need to operate on “facts, data and numbers”, but this can be difficult when these facts, data and numbers are highly error-prone. In the case of daily COVID-19 deaths in New York, there was serious disagreement even between sources regarding the ground truth for the number of deaths. A key take-away from our analysis is that \textbf{serious thought and investment must be made in quality data collection when it comes to COVID-19 daily death data, as well as COVID-19 resource utilisation} Clinical trial methodology (\cite{dave}) for data quality control must be brought to bear, especially when the consequences of policy decisions can so dramatically impact the lives of millions of people.

Early on, Dr. Anthony Fauci, NIAID Director, stated that \cite{bump}: ``As I've told you on the show, models are really only as good as the assumptions that you put into the model. But when you start to see real data, you can modify that model...” An open question raised by this thoughtful comment is how can one expect quality predictions, when the data are suspect? How does one modify the model in light of the data, if the data are faulty? Would the course of action of policy leaders have differed, had it been known that there would not be clear agreement on what represented ground truth for even a hard endpoint such as death?
Clearly, if the data are suspect, projections may also be sub-optimal. 

However, putting the issue of data quality aside, our analysis shows that models tended to have very poor performance both in terms of accuracy as well as in terms of capturing uncertainty. To be fair, pandemics are, thankfully, rare events and predicting outcomes in the early stages is very difficult, as so much is unknown. Rosen \cite{rosen} quotes Dr. Alain Labrique 
\begin{quoting}
\noindent
\textit{“With a new virus, and any type of infectious disease that we have never encountered before, there are many unknowns," said Dr. Alain Labrique, an associate professor at The Johns Hopkins Bloomberg School of Public Health and one of the nation's most renowned epidemiologists." And so the big challenge for us is to focus on telling the public the truth about what we know, and to explain the uncertainties around what we don't know."}
\end{quoting}

In this regard, the LANL model was the only model that was found to approach the 95\% nominal coverage, but unfortunately this model was unavailable at the time Governor Cuomo needed to make major policy decisions in late March 2020. Model predictions for daily deaths tended to have smaller errors over time, but this is not reassuring, because predictions are extremely critical in the early phase of an epidemic wave. 

The importance of accurate early predictions applies even more to predictions for bed utilisation, where wrong expectations can lead to wrong decisions. For example, a major mistake in New York was the decision to send COVID-19 patients to nursing homes. Based on a March 25 directive, over 4\,500 COVID-19 patients were discharged from hospitals to nursing homes (\cite {apnews}), specifically because it was anticipated that regular hospital beds would be urgently needed and hospitals would be overrun by COVID-19 patients. Nursing homes are full of highly vulnerable people and outbreaks in nursing homes (\cite {abrams}) resulted in high fatalities. In New York alone, over 5\,800 deaths occurred in nursing homes (\cite{apnews}). Eventually this was a sizeable fraction of the COVID-19 death burden, and importantly, it might have been avoidable to a large extent. Overestimates of anticipated bed requirements could also have affected hospital utilisation for other serious conditions with adverse consequences for the outcomes of patients suffering from these conditions (\cite{fil}, \cite{sud}). While preparedness is important and beneficial, making preparations with vastly erroneous expectations can create major harm.   

Our second key take-away from this evaluation: \textbf{the need for real time evaluation of prediction models}. Going forward there needs to be industry standards as to how models are to be evaluated and calibrated in real time, especially in the rapidly evolving settings of a pandemic.
Quoting Dr. B. Jewell: ``This appearance of certainty is seductive when the world is desperate to know what lies ahead" \cite{begley}. Unfortunately, in retrospect, COVID-19 anxiety can turn to COVID-19 disillusionment when the decisions made by policy makers are dictated by suspect low quality data and consequently by poorly performing models. One solution would be to compare predictions of models against emerging reality on a daily basis using prespecified metrics such as those analysed here. Models that are consistently poorly performing should carry less weight in shaping policy considerations.  Models may be revised in the process, trying to improve performance. However, improvement of performance against retrospective data offers no guarantee for continued improvement in future predictions. Failed and recast models should not be given much weight in decision making until they have achieved a prospective track record that can instil some trust for their accuracy.  Even then, real time evaluation should continue, since a model that performed well for a given period of time may fail to keep up under new circumstances.

\bibliographystyle{abbrv}
\bibliography{biblio}

\end{document}